\documentclass[sigconf]{acmart}

\renewcommand\footnotetextcopyrightpermission[1]{} 
\usepackage{algorithm}
\usepackage{algpseudocode}

\usepackage{sidecap}

\AtBeginDocument{%
  \providecommand\BibTeX{{%
    \normalfont B\kern-0.5em{\scshape i\kern-0.25em b}\kern-0.8em\TeX}}}

\setcopyright{acmcopyright}
\copyrightyear{2020}
\acmYear{2020}

\acmConference[SIGIR '20]{Proceedings of the 43rd International ACM SIGIR Conference on Research and Development in Information Retrieval (SIGIR '20)}{July 25-30, 2020}{Xi'an, China}

\begin{document}

\title{\textsc{Choppy}: Cut Transformer For Ranked List Truncation}

\author{Dara Bahri}
\affiliation{
\institution{Google Research}
}
\authornote{Corresponding author}
\email{dbahri@google.com}
\author{Yi Tay}
\affiliation{
\institution{Google Research}
}
\email{yitay@google.com}
\author{Che Zheng}
\affiliation{
\institution{Google Research}
}
\email{chezheng@google.com}
\author{Donald Metzler}
\affiliation{
\institution{Google Research}
}
\email{metzler@google.com}
\author{Andrew Tomkins}
\affiliation{
\institution{Google Research}
}
\email{tomkins@google.com}





\begin{abstract}
Work in information retrieval has traditionally focused on ranking and relevance: given a query, return some number of results ordered by relevance to the user. However, the problem of determining how \textit{many} results to return, i.e. how to optimally truncate the ranked result list, has received less attention despite being of critical importance in a range of applications. Such truncation is a balancing act between the overall relevance, or usefulness of the results, with the user cost of processing more results. In this work, we propose \textsc{Choppy}, an assumption-free model based on the widely successful Transformer architecture, to the ranked list truncation problem. Needing nothing more than the relevance scores of the results, the model uses a powerful multi-head attention mechanism to directly optimize any user-defined IR metric. We show \textsc{Choppy} improves upon recent state-of-the-art methods.
\end{abstract}




\maketitle

\def\reals{\mathbb{R}}
\def\expect{\mathbb{E}}

\section{Introduction}
While much of the work in information retrieval has been centered around ranking, there is growing interest in methods for \textit{ranked list truncation} - the problem of determining the appropriate cutoff $k$ of candidate results \cite{arampatzisSIGIR09,lienICTIR19}. This problem has garnered attention in fields like legal search \cite{tomlinsonTREC07} and sponsored search \cite{broderCIKM08,wangPAKDD11}, where there could be a monetary cost for users looking into an irrelevant tail of documents or where showing too many irrelevant ads could result in ad blindness. The fundamental importance of this problem has led to development of methods that are automatically able to learn $k$ in a data-driven fashion \cite{lienICTIR19}. The focus of this paper is to design more effective models for accurate and dynamic truncation of ranked lists.

 The present state-of-the-art for this task is BiCut \cite{lienICTIR19}, a recurrent-based neural model that formulates the problem as a sequential decision process over the list. BiCut trains a bidirectional LSTM \cite{hochreiter1997long} model with a predefined loss that serves as a proxy to the user-defined target evaluation metric. At every position in the ranked list, BiCut makes a binary decision conditioned on both forward and backward context: to continue to the next position, or to end the output list.

While BiCut outperforms non-neural methods \cite{arampatzisSIGIR09}, we argue it has several drawbacks. Firstly, the model is trained with teacher-forcing, i.e. with ground truth context, but it is deployed auto-regressively at test time, where it is conditioned on its own predictions. Thus, the model suffers from a train / test distribution mismatch, often referred to as exposure bias \cite{ranzato2015sequence}, resulting in poor model generalization. Secondly, the loss function used does not capture the mutual exclusivity among the candidate cut positions. In other words, the loss does not capture the condition that the list can only be cut in at most one position. Furthermore, the proposed training loss is unaligned with the user-defined evaluation metric. Last but not least, BiCut employs BiLSTMs which are not only slow and non-parallelizable, but also do not take into account global long-range dependencies. 

This paper proposes \textsc{Choppy}, a new method that not only ameliorates the limitations of the BiCut model but also achieves state-of-the-art performance on the ranked list truncation task. Our method comprises two core technical contributions. The first is a Transformer model \cite{vaswani2017attention} that is able to capture long-range dyadic interactions between relevance scores. To the best of our knowledge, this is the first successful application of self-attentive models on scalar ranking scores. The second technical contribution is the development of a loss function that optimizes the \textit{expected} metric value over all candidate cut positions. Overall, \textsc{Choppy} not only improves the predictive performance on this task but also improves the model inference speed by $>3$ times.

\paragraph{Our Contributions}
The key contributions of the paper are summarized as follows:
\begin{itemize}
    \item We frame the ranked list truncation task as modeling the joint distribution among all candidate cut positions, and we construct our training loss to be the \textit{expected} metric value over cut positions, for \textit{any} choice of user-defined metric, such as F1, precision, or discounted cumulative gain. As such, our training loss and evaluation metric are fully aligned.
    \item We propose \textsc{Choppy}, a Cut Transformer model that achieves a state-of-the-art $11.5\%$ relative improvement over the BiCut model. To predict the joint distribution over candidate cut positions, our method learns a positional embedding and leverages expressive bidirectional multi-headed attention.
\end{itemize}



\begin{figure}[ht!]
\centering
\includegraphics[width=0.33\textwidth]{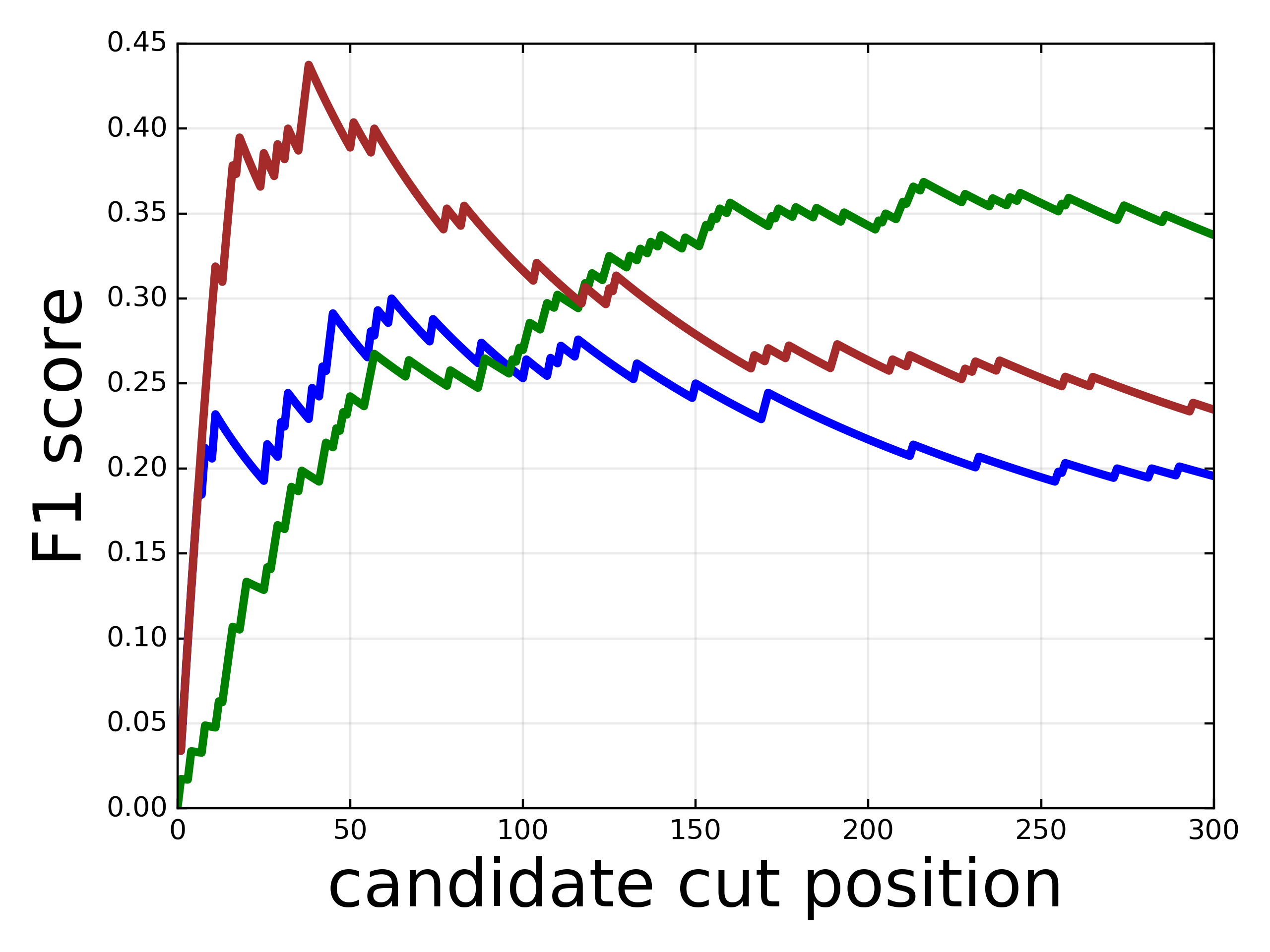}\\ \includegraphics[width=0.33\textwidth]{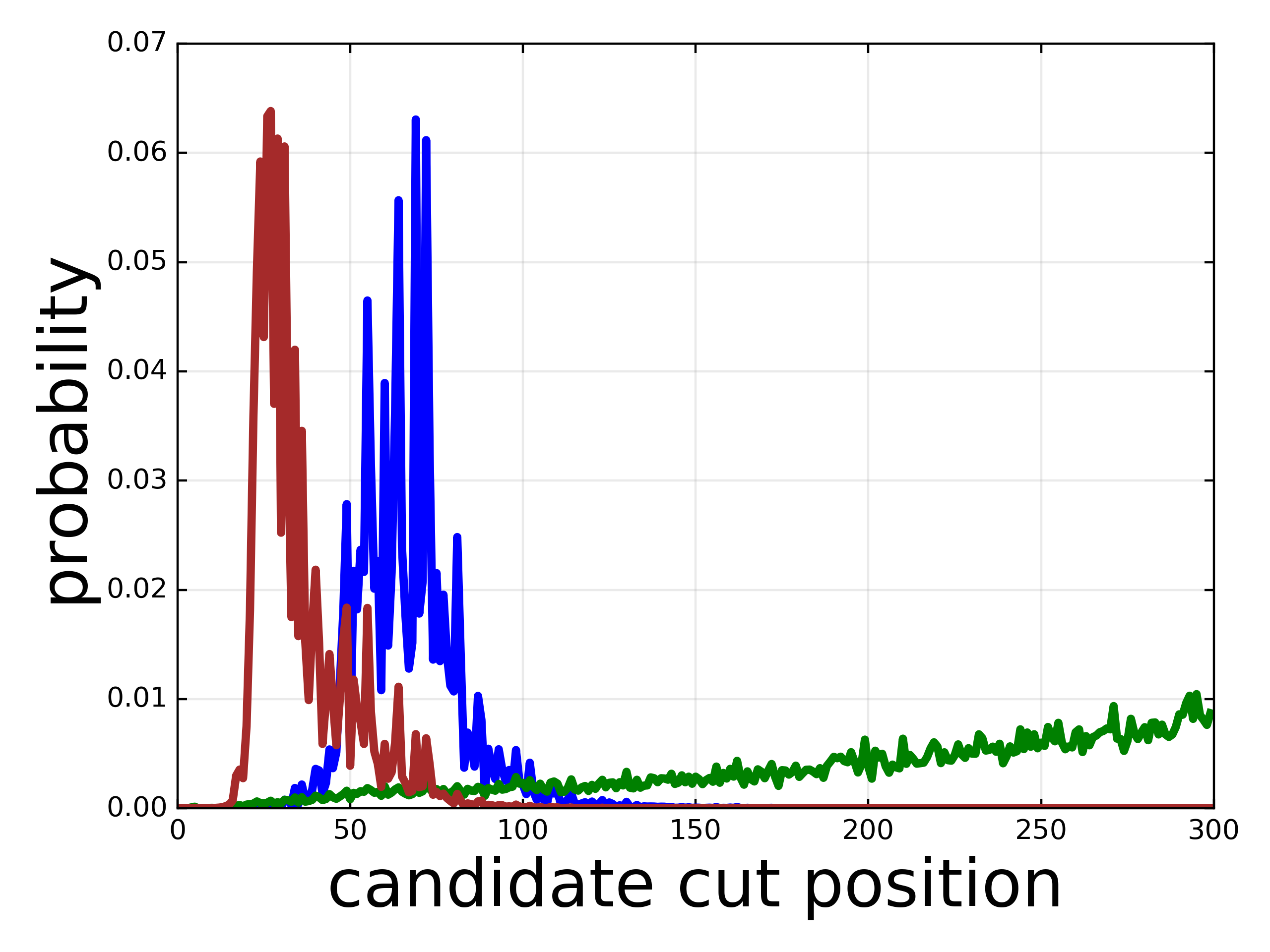}
\caption{Top: F1 at various cut positions for 3 training queries from Robust04 BM25. Bottom: \textsc{Choppy}'s softmax predictions for the same queries.}
\label{fig:loss}
\end{figure}


\section{Related Work}\label{sec:related}
Across the rich history of information retrieval research, there has been extensive work focused on modeling score distributions of IR systems. Early work in this area primarily focused on fitting parametric probability distributions to score distributions \cite{manmathaSIGIR01,arampatzisSIGIR09}. This is often achieved by making the assumption that the overall distribution can be expressed as a mixture of a relevant and a non-relevant distribution. The expectation-maximization (EM) algorithm is often adopted to learn the parameters.

There has been considerable recent interest in adopting machine learning based models to optimize and improve the ranked list truncation problem. For instance, cascade-style IR systems \cite{wangSIGIR11} seek to achieve a balance between efficiency and effectiveness. Notably, \cite{culpepperADCS16} investigates a number of machine learning approaches for learning dynamic cutoffs within cascade-style ranking systems. Another recent study investigated how to leverage bidirectional Long Short-Term Memory (LSTM) models to identify the best position to truncate a given list \cite{lienICTIR19}. This model, BiCut, can be considered the present state-of-the-art approach.

Our work is closely related to the task of query performance prediction \cite{cronentownsendSIGIR02}. In this task, the objective is to automatically determine the effectiveness of a given query. This could be leveraged to determine the optimal set of results to the user for any given measure. Methods for query performance prediction include pre-retrieval-based approaches \cite{hauffCIKM08}, relevance-based approaches \cite{cronentownsendSIGIR02,zhouSIGIR07}, and neural approaches \cite{zamaniSIGIR18}.

A system that determines the best number of results to display to users has the potential to benefit a wide number of applications. For example, in sponsored search, displaying too many irrelevant ads to users may cause frustration, resulting in so-called \textit{query blindness}. This motivated research that investigated whether any ads should be displayed at all \cite{broderCIKM08}. It is also easy to see that a similar and related problem formulation is to determine how many ads should be displayed to the users \cite{wangPAKDD11}. Moreover, determining the optimal number of ranked results is also important in a number of other IR applications such as legal e-discovery  \cite{tomlinsonTREC07}, where there is an significant financial or labor cost associated with reviewing results. Finally, the ability to calibrate scores across queries and different corpora has also been studied in the context of federated search tasks \cite{shokouhiFTIR11} such as meta-search \cite{montagueCIKM01}.

\section{\textsc{Choppy}}\label{sec:choppy}
We now describe \textsc{Choppy}, our proposed Cut Transformer approach for ranked list truncation.

Let $(\mathbf{r}_1,\dots,\mathbf{r}_n)$ denote the sequence of results, ranked in decreasing order of relevance, and let $\mathbf{r}_i$ have relevance score $\mathbf{s}_i$ and ground truth relevance label $\mathbf{y}_i$ ($1$ if relevant, $-1$ if non-relevant). We now describe our model piecemeal.

\subsection{Transformer Layer}
We briefly review the Transformer layer. In our work, we let all model dimensions be $d$.
Let $X \in \reals^{n \times d}$ represent the input to the layer and $W_q, W_k, W_v \in \reals^{d \times d}$.
We define
\begin{align*}
    \operatorname{Attn}\left(X; W_q, W_k, W_v\right) = \operatorname{Act}\left(XW_q\,W_k^TX^T\right)\left(XW_v\right)
\end{align*}
where $\text{Act}$ is an activation function. As done in \cite{vaswani2017attention}, we take it to perform row-wise softmax and normalize by $\sqrt{d}$. Attention is often augmented using multiple heads. In multi-headed attention, the model dimension is split into multiple heads, each one performing attention independently, and the result is concatenated together. Let $h$ be the number of heads and suppose $d$ is divisible by $h$. Then,
\begin{align*}
    \operatorname{MultiAttn}\left(X; W_q, W_k, W_v\right) = \operatornamewithlimits{rConcat}\limits_{1 \leq i \leq h}\left[\operatorname{Attn}\left(X; W_q^{(i)}, W_k^{(i)}, W_v^{(i)}\right)\right]
\end{align*}
where $W_q^{(i)},W_k^{(i)},W_v^{(i)} \in \reals^{d \times (d/h)}$. $\text{rConcat}$ performs row-wise concatenation.
The output of multi-headed attention is
\begin{align*}
    A = \operatorname{LayerNorm}\left(X + \operatorname{MultiAttn}\left(X; W_q, W_k, W_v\right)\right)
\end{align*}
where $\operatorname{LayerNorm}$ is layer normalization \cite{ba2016layer}.
Finally, the output of the Transformer layer is
\begin{align*}
    O_\text{trans} = \operatorname{LayerNorm}\left(A + \operatorname{rFF}\left(A\right)\right)
\end{align*}
where $\operatorname{rFF}$ applies a single learnable feed-forward layer with ReLU activation to each row of $A$.

\subsection{Positional Embedding}
The vanilla Transformer incorporates positional encoding by adding fixed sinusoidal values to the input token embeddings. As the token embeddings are trained, they have the flexibility to learn how to best utilize the fixed positional information. In our setting however, the inputs are fixed 1-dimensional relevance scores. Attempting to apply a Transformer layer directly on the raw scores can limit its complexity. To that end, we introduce a learnable positional embedding $P \in \reals^{n \times (d - 1)}$ and feed in $X = \operatorname{rConcat}\left[\mathbf{s}, P\right]$
 to the the first Transformer layer only, where $\mathbf{s}$ is the column vector of relevance scores.

\subsection{Loss}
So far, Choppy takes the $n$-length vector of scores, augments it with a positional embedding, and feeds the result into $n_\text{layers}$ Transformer layers. This produces $O_\text{trans} \in \reals^{n \times d}$. We arrive at the output of Choppy by applying a final linear projection followed by a softmax over positions:
\begin{align*}
\mathbf{o} = \operatorname{Softmax}\left(O_\text{trans} \,W_o\right)
\end{align*}
where $W_o \in \reals^{d \times 1}$. We interpret the output $\mathbf{o}$ to be a probability distribution over candidate cutoff positions. More concretely, we take $\mathbf{o}_i = \operatorname{Prob}\left[\left(\mathbf{r}_1, \dots, \mathbf{r}_i\right)\right]$. Let $C$ be any user-defined evaluation metric that should be maximized, such as F1 or precision. For each training example $j$ and every candidate cutoff position $i$ we compute $C_i(\mathbf{y}^{(j)})$, the value of the metric if the result list were to be truncated at position $i$, using the ground-truth relevance labels. Our proposed loss follows as:
\begin{align*}
    L\left(\mathbf{s}^{(j)}, \mathbf{y}^{(j)}\right) &= -\sum_{i=1}^{n} \mathbf{o}_i\left(\mathbf{s}^{(j)}\right) C_i\left(\mathbf{y}^{(j)}\right)\\
    &= -\expect_{Z \sim \operatorname{Categorical}\left(\mathbf{o}\left(\mathbf{s}^{(j)}\right)\right)}\,C_Z\left(\mathbf{y}^{(j)}\right).
\end{align*}
With this loss, our model learns the conditional joint distribution over candidate cut positions that maximizes the expected evaluation metric on the training samples. We depict the loss and the predicted distribution for a few training samples in Figure~\ref{fig:loss}. We see that the model tends to weight positions according to their corresponding metric value. At test time we choose to cut at the argmax position. Note that unlike BiCut, our loss has no tune-able hyperparameters.
\begin{figure*}[!ht]
\begin{tabular}{cc}
  \includegraphics[width=0.35\textwidth]{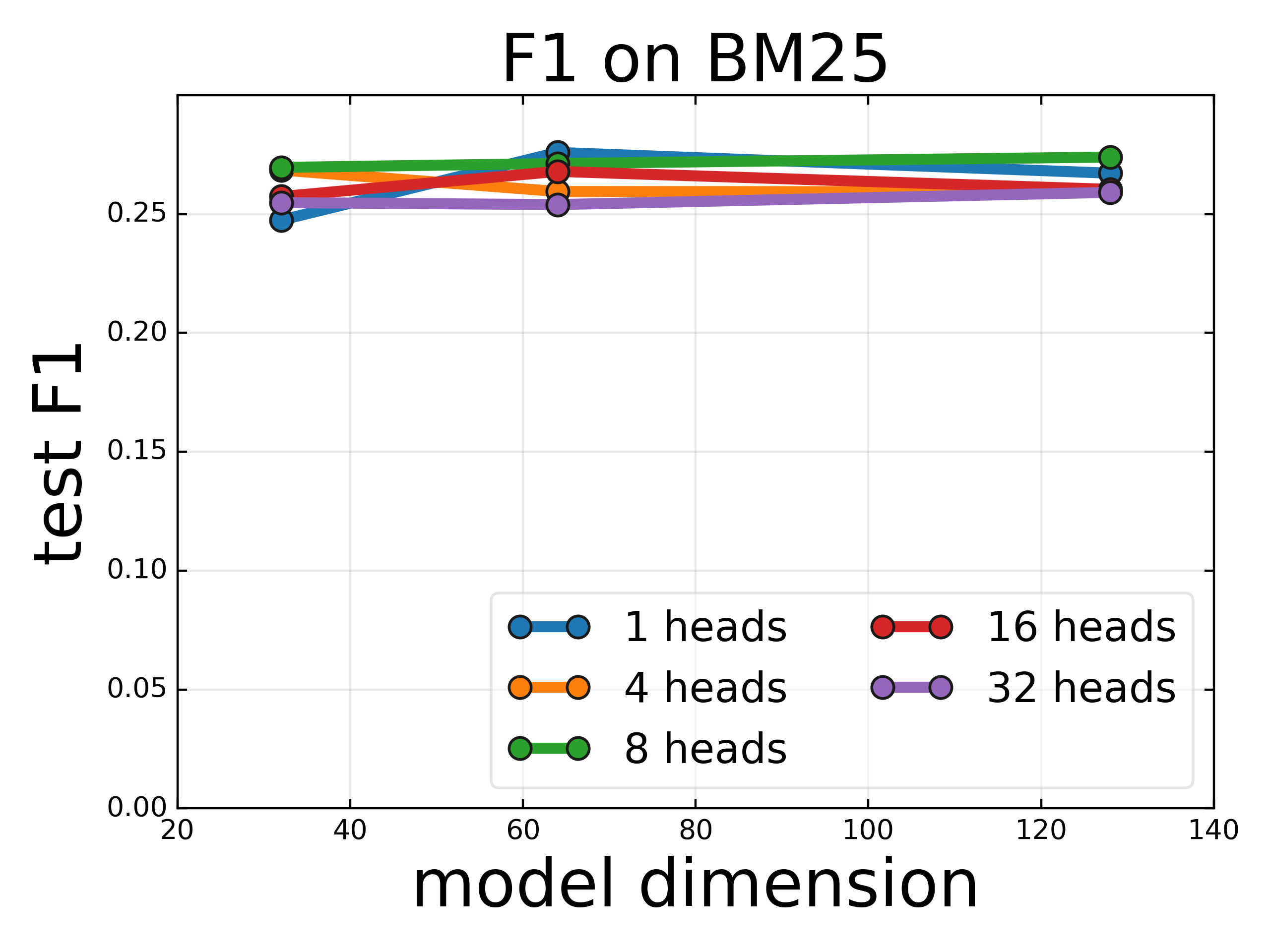} & \includegraphics[width=0.35\textwidth]{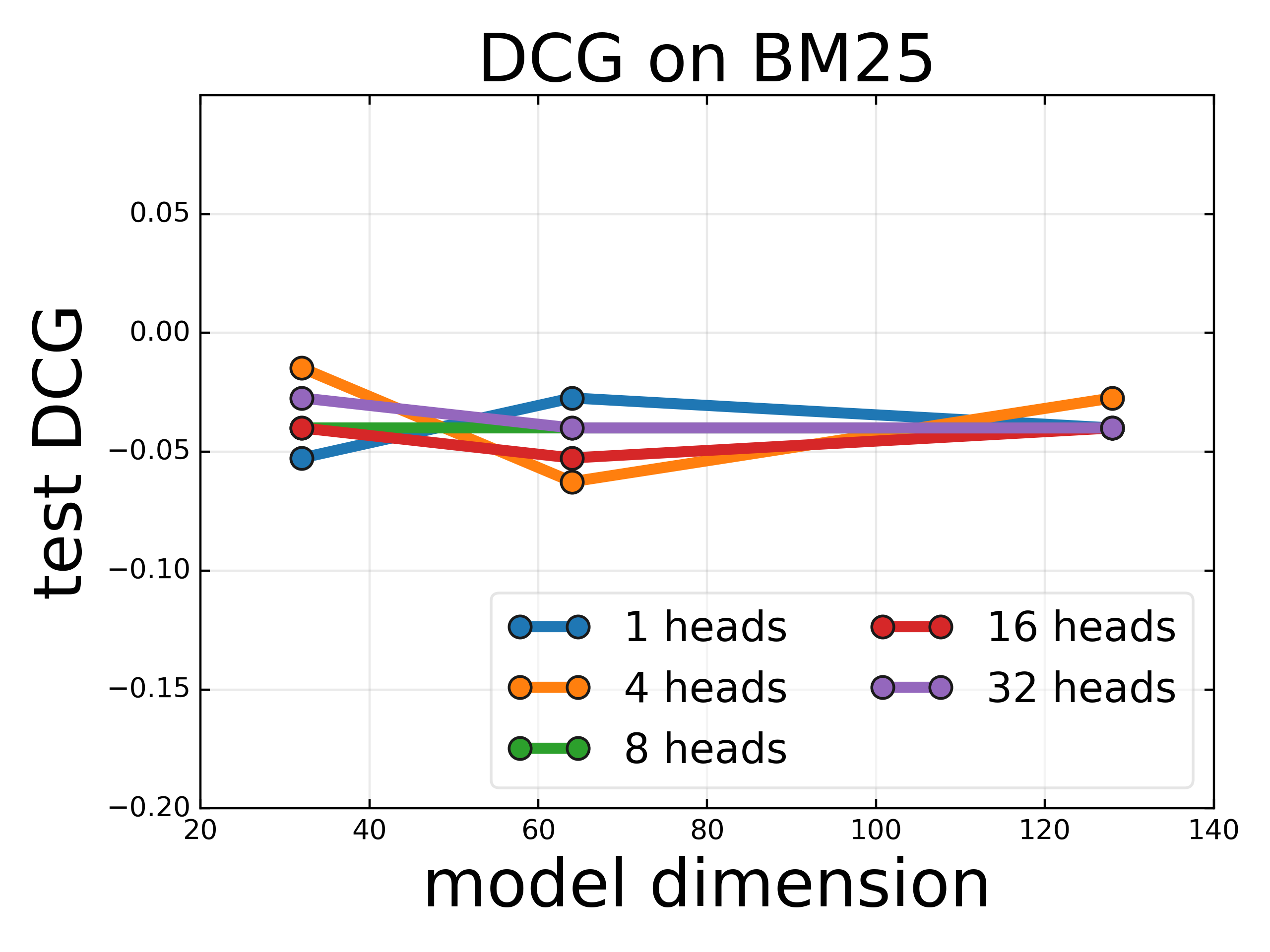} 
\end{tabular}
\caption{Performance on Robust04 BM25 for different model dimensions $d$ and heads $h$ ($n_\text{layers} = 3$). We see strong and stable performance across different settings. Note that the y-axis is zoomed in to show details.}
\label{fig:ablation_bm25}
\end{figure*}

\section{Experiments}\label{sec:evaluation}
This section describes our experimental setup and results.
\subsection{Dataset}
We evaluate our method using the TREC collection Robust04, used in the TREC 2004 Robust Track. It consists of 250 queries over 528k news articles, where each query has 1000 total results and an average of 70 relevant
ones. This is the same dataset used in \cite{lienICTIR19}. We use a random 80/20 train/test that achieves comparable performance to the reported results in \cite{lienICTIR19}. We evaluate the efficacy of our truncation model using two different retrieval approaches - BM25, a traditional tf-idf based model, and DRMM \cite{guo2016deep}, a neural model.

\subsection{Baselines}
We evaluate our method against the following baselines:
\begin{itemize}
\item \textbf{Fixed-$k$} returns the top-$k$ results for a single value of $k$ across test queries.
\item \textbf{Greedy-$k$} chooses the single $k$ that maximizes $C$ over the training set.
\item \textbf{Oracle} uses knowledge of each test query's true label to optimize $k$. It represents an upper-bound on the metric performance that can be achieved.
\item \textbf{BiCut} \cite{lienICTIR19} learns a multi-layer bidirectional LSTM model on the entire training set, taking the score sequence as inputs. At position $i$ of the result list, the model predicts probability $p_{i}$ to continue and probability $1-p_i$ to \textbf{end}. At inference time, the cutoff is made before the first occurrence of \textbf{end}.
\end{itemize}

\subsection{Setting}
We report F1 and Discounted Cumulative Gain (DCG) scores where we define DCG to penalize negative, or non-relevant results:
\begin{align*}
\text{DCG}_n = \sum_{i=1}^n \frac{y_i}{\log_2 (i+1)},
\end{align*}
We need to deviate from the usual definition of DCG since the usual definition always increases monotonically with the length of the returned ranked list and so the optimal solution under this definition would be to not truncate at all. For methods that optimize F1 or DCG, we report the performance of the model when it is optimized specifically for that metric. Note that DCG is unsupported by BiCut.

For \textsc{Choppy}, we blithely set $n_\text{layers} = 3$, $h\, (\text{\# heads}) = 8$, and $d = 128$ across all settings, without any tuning. We optimize the aforementioned custom loss function using Adam with default learning rate $0.001$, and a batch size of $64$. As in \cite{lienICTIR19}, we only consider the top-$300$ candidate results of each query. 

\begin{table}[]
\small
\begin{tabular}{|r|cc|cc|}
\hline
& \multicolumn{2}{c|}{BM25}                           & \multicolumn{2}{c|}{DRMM}                          \\ \hline
                      & F1             & DCG             & F1             & DCG            \\ \hline
Oracle                & 0.367          & 1.176           & 0.375          & 1.292          \\
\hline
Fixed-$k$ (5)         & 0.158          & -0.261          & 0.151          & 0.010          \\
Fixed-$k$ (10)        & 0.209          & -0.708          & 0.197          & -0.407         \\
Fixed-$k$ (50)        & 0.239          & -5.807          & 0.261          & -5.153         \\
Greedy-$k$            & 0.248          & -0.116          & 0.263          & 0.266          \\
BiCut                 & 0.244          & -               & 0.262          & -              \\
\hline
\textsc{Choppy}                & \textbf{0.272} & \textbf{-0.041} & \textbf{0.268} & \textbf{0.295} \\
Rel. \% Gain & +11.5\% & - & +2.29\% & - \\
\hline
\end{tabular}
\caption{Average F1 and DCG performance on Robust04. Choppy achieves state-of-the-art performance. ``Gain'' reports relative performance gain over BiCut model. }\label{tab:robust04_results}
\end{table}

\section{Results and Discussion}
\subsection{Results}
As shown in Table~\ref{tab:robust04_results}, \textsc{Choppy} achieves a significant improvement over BiCut for both metrics and both retrieval types. This improvement arises from \textsc{Choppy}'s ability to model the joint distribution over \textit{all} candidate cut positions and its direct optimization of the evaluation metric. Furthermore, the attention mechanism is able to effectively capture correlations between scores far apart in ranked order. This is in contrast to LSTMs, as used in BiCut, whose degradation with larger sequence length is well known.

\subsection{Ablation Study}
Choppy has three hyperparameters: the model dimension $d$, the number of heads $h$, and the number of Transformer layers $n_\text{layers}$. In Figure~\ref{fig:ablation_bm25} we plot the impact of $d$ and $h$ on predictive performance (while keeping $n_\text{layers}$ fixed to $3$). We see that both F1 and DCG are strong and stable across settings. Extrapolating beyond Robust04, we expect \textsc{Choppy} to work out-of-the-box on many datasets. Unlike BiCut, which required much tuning in our experience, \textsc{Choppy} seems to work well while requiring little-to-no tuning.

\section{Conclusion}\label{sec:conclusions}
We propose \textsc{Choppy}, a Transformer architecture for ranked list truncation, that learns the score distribution of relevant and non-relevant documents and is able to directly optimize \textit{any} user-defined metric. We show that \textsc{Choppy} achieves state-of-the-art F1 and DCG performance on Robust04, under both a traditional tf-idf as well as modern neural ranking system. We then dig deeper into \textsc{Choppy}'s architecture settings, showing strong and stable performance across a range of values. We thus conclude that when faced with a ranked list truncation task, one can apply \textsc{Choppy} and expect competitive performance.

\begin{acks}
We would like to thank the authors of BiCut for providing us with the BM25 and DRMM ranked lists that were used for evaluation.
\end{acks}

\bibliographystyle{ACM-Reference-Format}
\bibliography{references}


\end{document}